\documentclass[openacc]{rstransa}

\usepackage{soul}

\newcommand{\permpc}{Mpc$^{-1}$}
\newcommand{\kpar}{$k_{\parallel}$}
\newcommand{\kperp}{$k_{\perp}$}
\newcommand{\sarc}{$''$}

\begin{document}

\title{Reaching small scales with low frequency imaging: applications to the Dark Ages}

\author{
L. K. Morabito$^{1}$ and J. Silk$^{2,3}$ }

\address{$^{1}$Centre for Extragalactic Astronomy, Department of Physics, Durham University, Durham, DH1 3LE, UK\\
$^{2}$Institut  d'Astrophysique  de  Paris,  UMR  7095  CNRS,Sorbonne  Universit\'{e}s,  98  bis,  boulevard  Arago,  F-75014,  Paris,  France \\
$^{3}$Beecroft  Institute  of  Particle  Astrophysics  and  Cosmology,  University  of  Oxford,  Oxford  OX1  3RH,  UK \\ }

\subject{observational astronomy, cosmology}


\corres{Leah Morabito\\
\email{leah.k.morabito@durham.ac.uk}}

\begin{abstract}
The initial conditions for the density perturbations in the early Universe, which dictate the large scale structure and distribution of galaxies we see today, are set during inflation. Measurements of primordial non-Gaussianity are crucial for distinguishing between different inflationary models. Current measurements of the matter power spectrum from the CMB only constrain this on scales up to $k\sim 0.1\,$\permpc . Reaching smaller angular scales (higher values of $k$) can provide new constraints on non-Gaussianity. A powerful way to do this is by measuring the $\textrm{H}\textrm{I}$ matter power spectrum at $z\gtrsim30$. In this paper, we investigate what values of $k$ can be reached for the LOw Frequency ARray (LOFAR), which can achieve $\lesssim$1\sarc $\,$resolution at $\sim$50 MHz. Combining this with a technique to isolate the spectrally smooth foregrounds to a wedge in \kpar - \kperp\ space, we demonstrate what values of $k$ we can feasibly reach within observational constraints. We find that LOFAR is $\sim$5 orders of magnitude away from the desired sensitivity, for 10 years of integration time. 
\end{abstract}

\begin{fmtext}
\section{Introduction}

The end of Inflation set the density fluctuations which dictated the growth of large-scale structure and the distribution of galaxies we see today. Observations of the Cosmic Microwave Background (CMB) and galaxy surveys which trace the large scale structure have been used to constrain the matter power spectrum on large scales (co-moving wavenumbers $k\lesssim 0.1\,$\permpc \cite{planck_arxiv,gil_marin}) while observations of galaxy clustering, weak gravitational lensing and the Lyman-$\alpha$ forest extend this  {\color{white}{of}}
\end{fmtext}

\maketitle


\noindent to the range of $ 0.1 \lesssim k \lesssim 1\,$\permpc\ (e.g., Fig 19 of \cite{planck1}). Probing higher co-moving wavenumbers provides new information, and places powerful constraints on the distribution of matter  in the early Universe. Using the spin-flip transition of neutral hydrogen allows us to access the matter distribution in the Dark Ages (after the formation of the first stars, but before reionisation). At high enough redshifts, the temperature of the Universe was coupled with the CMB temperature through Compton scattering, but after $z\sim 200$ the Universe began to cool more rapidly than the CMB. This provides the ideal conditions for detecting the spin-flip in absorption. 
At low redshifts, the small scales (high $k$) are heavily influenced by the complex astrophysical processes of galaxy formation and reionization. Reaching above $z\sim 30$ allows us to avoid many of these complications. 

Measuring the 21 cm power spectrum in the early Universe is the goal of many low-frequency instruments, which usually focus on the lower redshifts of the Epoch of Reionisation  \cite{neben,parsons_eight,vh,monsalve,beardsley,mertens,shaw}. Some experiments focus on achieving a global detection, which measures the average temperature brightness across the sky of the 21 cm line, $T_{21}$, as a function of redshift. This is an inexpensive way to make a detection, as in principle it can be done with a single antenna and enough integration time. However, astrophysical foregrounds have to be removed or simultaneously fitted for in the data to extract the 21 cm signal, and it can be hard to interpret the results (seen recently with the claimed detection from EDGES \cite{bowman}).

Although more resource intensive, using interferometers to measure 21 cm fluctuations can provide more stringent constraints. With resolution comes the possibility of isolating foregrounds to a `wedge' in $k$ space \cite{datta,vedantham,trott,pober21,liu}. In the 2D power spectrum, foregrounds will occupy low values of \kpar\ (i.e., along the line of sight). The imperfect sampling of smoothly varying foreground spectra results in an upscattering of the foreground power in this parameter space, but the foregrounds can still be isolated to a particular region. By cleanly isolating the foregrounds, we can make the most of the data which probes the rest of the \kpar - \kperp\ parameter space. 

Angular resolution offers more than just foreground isolation: it allows us to probe higher values of \kperp , which is crucial for extracting new information from the data to place constraints on the 21 cm power spectrum \cite{cole,munoz}. By measuring the small-scale matter distribution in the Dark Ages, we can place completely new constraints on models of inflation, and simultaneously help probe open questions on the more nearby Universe. For example, measuring the small-scale power spectrum can inform us on whether or not primordial black holes were large enough to explain the rapid growth of the most massive super-massive black holes at $z \gtrsim 6$ \cite{carr,latif}. It is also possible that primordial black holes can increase our chances of detecting the 21 cm signal: their presence can deepen the absorption at $z\lesssim 50$ when both the product of Eddington ratio and mass, as well as their source density, are large \cite{bernal}. This enhancement in the absorption is expected to happen at $k \gtrsim 10^{-2}$-$10^{-1}\,$\permpc , and increase for larger values of $k$. 

To reach the relevant values of \kperp , it is necessary to use an interferometer with high resolution and a wide field of view. The LOw Frequency ARray (LOFAR) \cite{vh} is the only low-frequency radio interferometer capable of the necessary sub-arcsecond resolution, and simultaneously offers a 5 deg$^2$ field of view. With stations currently spread across 7 countries (the Netherlands, Germany, UK, France, Sweden, Ireland, Poland) and growing (Latvia and Italy are building stations, and there are future plans for expansion to other countries), its capabilities will remain unique even in the era of the Square Kilometre Array (SKA). As such, it is interesting to ask the question whether LOFAR could be used to detect the small-scale 21 cm power spectrum, which is the focus of this speculative study. We intend this to be informative only, with an eye towards the future construction of a low-frequency lunar array. Such an array would remove the ionosphere (opening up the regime below $\sim$10 MHz, where the ionosphere becomes opaque) and mitigate radio frequency interference \cite{jester}. Currently LOFAR's longest baseline is about two-thirds that of the diameter of the moon, and offers a real-world example of the type of long-baseline low-frequency array that could be built. 

The paper is organised as follows: in Section 2 we discuss using LOFAR as a cosmological instrument, with particular attention to its high-resolution imaging capability. In Section 3 we address foreground isolation. Section 4 presents our sensitivity predictions, with discussion in Section 5 and conclusions and future work in Section 6. 

Throughout this paper we have assumed cosmological parameters from the \emph{Planck} 2013 results \cite{ade}.

\section{The Low Frequency Array as a cosmological instrument}

The LOw Frequency ARray (LOFAR) is a phased array comprised of fixed dipoles, with phase delays introduced into the signal paths to electronically `point' the telescope. Groups of 96 dipoles form stations, which are correlated in the same way as a traditional interferometer. LOFAR has two frequency bands: the Low Band Antenna (LBA) which operates between 10 -- 90 MHz, and the High Band Antenna (HBA) which operates between 110 and 240 MHz. There is a hard limit of 96 MHz of available bandwidth, which is typically split into 2$\times$48 MHz to conduct two simultaneous observations. In practice, this bandwidth is adjusted be centred on the sharply peaked maximum sensitivity of the LBA at $\sim$55 MHz, or to avoid the strong bandpass roll-off below 120 MHz for the HBA. The bandwidth is divided into 244 subbands, which can each be divided further into channels. The finest frequency resolution possible in a standard observing mode is 0.76 kHz (256 channels per subband), which corresponds to higher values of $k_{\parallel}$.

While default LOFAR operations use only the the stations in the Netherlands, baselines up to 1989 km (Ireland to Poland) provide sub-arcsecond imaging at MHz frequencies. This unique capability is competitive with higher-frequency instruments like \emph{e}-MERLIN, as LOFAR can achieve sub-arcseocnd imaging across a $\sim$5 deg$^2$ field of view in a single pointing. This high-resolution capability has been exploited to achieve scientific results on individual objects for both the HBA \cite{varenius} and LBA \cite{morabito}. The combination of high resolution and wide field of view also makes LOFAR an attractive survey instrument, and it is the \textit{only} instrument current capable of sub-arcsecond resolution at MHz frequencies, where the Dark Ages signal can be reached. This provides access to higher values of $k_{\perp}$. 

\begin{figure}[!h]
\centering\includegraphics[width=2.5in]{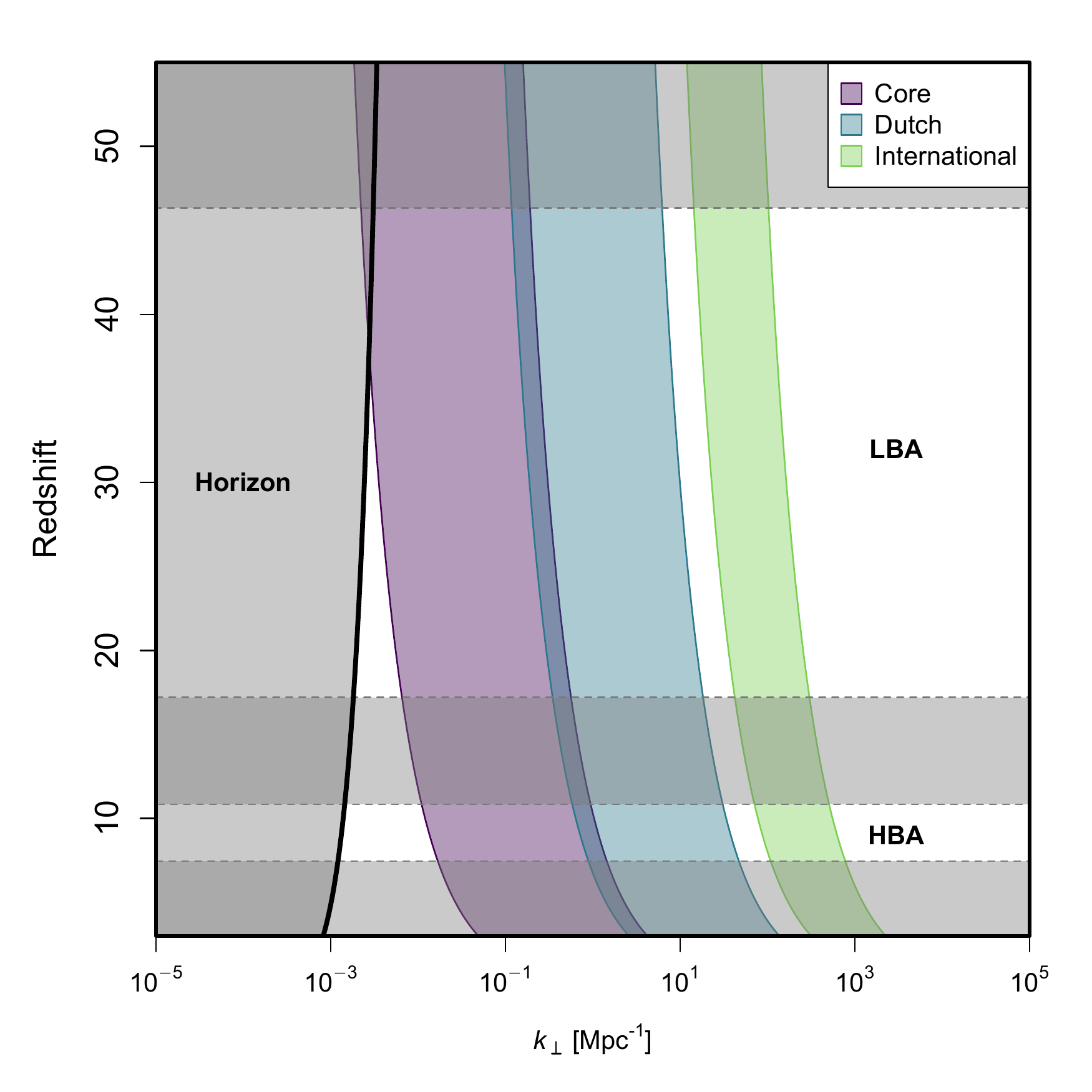}
\centering\includegraphics[width=2.5in]{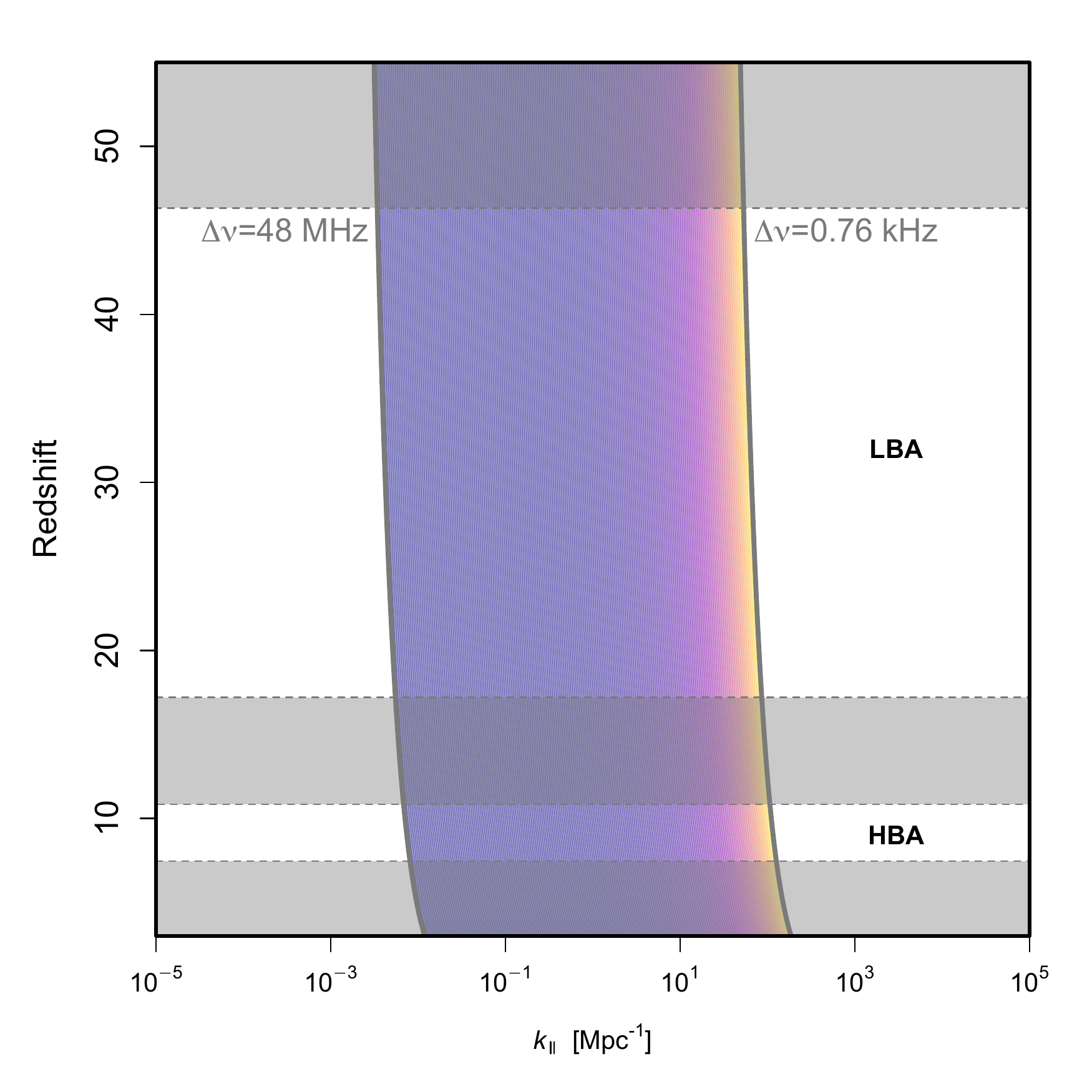}
\caption{Translating LOFAR's angular and spectral resolution into $k_{\perp}$ and $k_{\parallel}$. \textit{Left:} The shaded regions represent the values of $k_{\perp}$ which can be reached by the Dutch (blue) and International (green) stations. The gap in between them arises from the fact that there is a large geographical distance between the outermost Dutch stations and the closest international stations. The gray shaded region on the left shows the horizon limit. \textit{Right:} The gray lines show the limits on $k_{\parallel}$ based on the total bandwidth (48 MHz) and the smallest frequency resolution possible in a standard observing mode (0.76 kHz). The frequency spacing will be channelised, and the colour scale runs from wider channels (blue), which means more infrequent coverage in \kpar\ space, to to narrower channels (orange), which means more frequent coverage in \kpar\ space. This is plotted as a continuous shaded region, but in practice will be discrete. }
\label{fig1}
\end{figure}

Figure~\ref{fig1} shows what values of $k_{\perp}$ and $k_{\parallel}$ are currently available to LOFAR. Horizontal shaded regions indicate the redshift ranges which LOFAR cannot access because of its frequency coverage. This is based on the default observational setup and current array configuration. This includes 13 international stations, which provide $k_{\perp} \sim 10^{ 2.0 }$ - $10^{ 2.8 }\,$\permpc\  for the HBA and $k_{\perp} \sim 10^{ 1.5 }$ - $10^{ 2.3 }\,$\permpc\ for the LBA. The default frequency setup is 30 -- 78 MHz for the LBA and 120 -- 168 MHz for the HBA. The LOFAR EoR project \cite{patil,koopmans} uses a slightly different frequency range to access higher redshifts, but has to deal with the increasing radio frequency interference (RFI) at the top of the HBA band (the RFI rapidly increases above $\sim$168 MHz). The LBA is clearly more interesting for studying the Dark Ages, and in fact \cite{gehlot} have placed the first limits on the power spectrum for $z=19.8 - 25.2$. They use only the Dutch array, which limits \kperp\ to $<0.16\,$\permpc , and places a 2$\sigma$ limit of $\Delta^2_{21} < (\sim 1500\,\textrm{mK})^2$ at $k=0.038\,$\permpc\ from two simultaneously observed fields. They centre their bandwidth on the most sensitive portion of the LBA band, setting frequency limits of 39 - 72 MHz. Increasing the redshift upper limit to $z>100$ is technically possible by shifting the usable bandwidth down to start at the hard limit of 10 MHz, but this comes at the cost of greatly reduced sensitivity. It is worth noting that finer frequency increments provide diminishing returns in reaching higher values of $k_{\parallel}$. The limits are $k_{\parallel} \sim 10^{ -2.1 }$ - $10^{ 2.7 }\,$\permpc\ for the HBA and $k_{\parallel} \sim 10^{ -2.4 }$ - $10^{ 2.4 }\,$\permpc\ for the LBA. These calculations assume that there are no foregrounds to remove, which is addressed in the next section.

\section{Foregrounds}

Although LOFAR can theoretically access high-$k$ modes, this will be limited by foreground contamination. Setting aside the problem of calibrating for ionospheric distortions as the radio waves pass through the atmosphere (which would not be a problem for a lunar array), the foregrounds consist of Galactic emission and extragalactic sources. The Galactic foreground in particular has a steep power law spectrum, with $S_{\nu} \propto \nu^{-2.55}$. Extragalactic sources are also contaminants, and can range in shape and size from compact galaxies up to bright radio relics in galaxy clusters. Modelling and subtracting these forergrounds is perhaps the most obvious approach, but in practice this becomes extremely complex and difficult to do with high accuracy.

\begin{figure}[!h]
    \centering
    \includegraphics[width=4in]{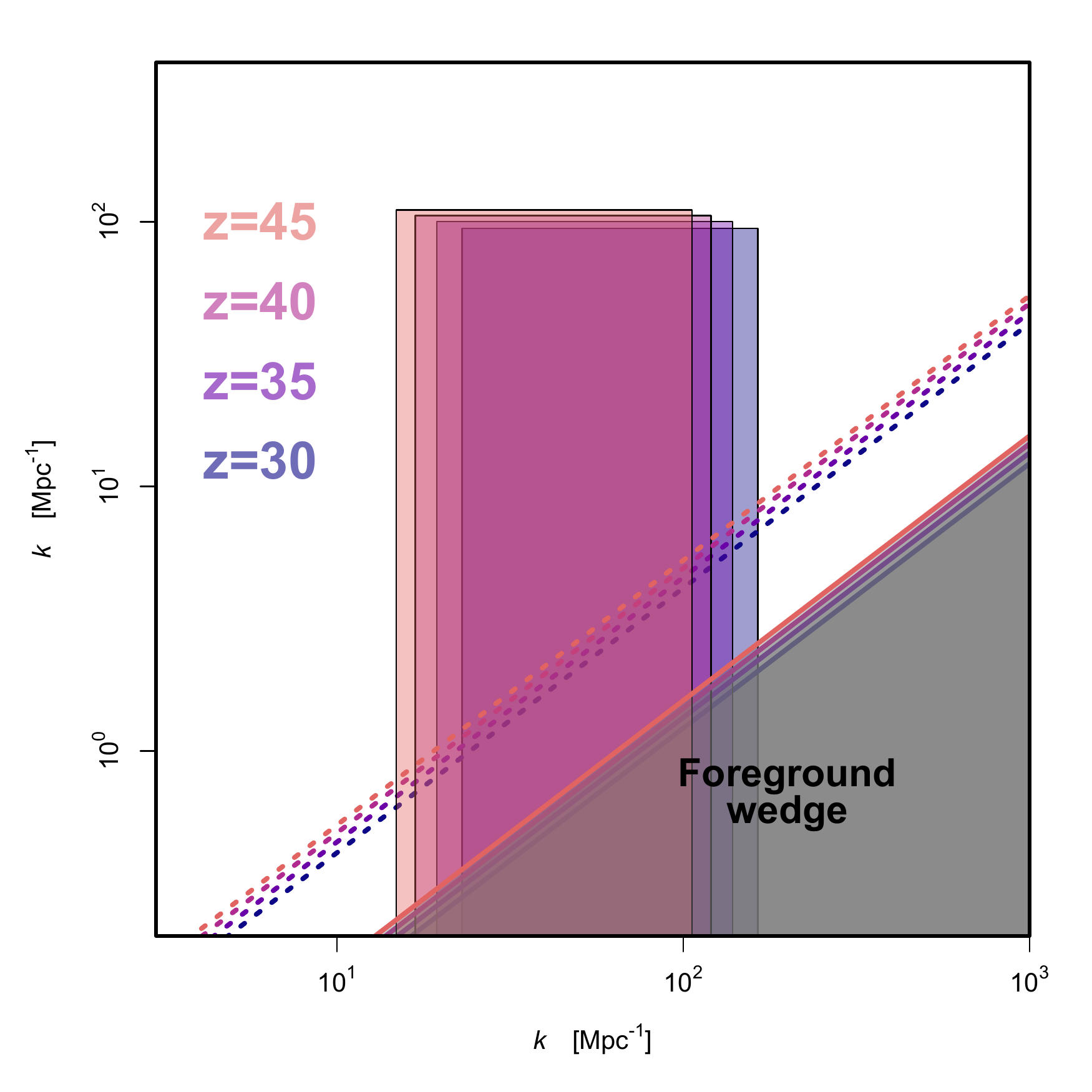}
    \caption{The shaded rectangles show the portion of the \kperp - \kpar\ plane which the angular and spectral resolution of LOFAR can access using the international stations, for four representative redshifts. The theoretical foreground wedge is shown as the gray shaded region in the bottom right hand corner of the plot. The dotted lines show the theoretical foreground wedge assuming the full LOFAR field of view using just the Dutch stations. The coloured rectangles and lines are for $z=45$ (light) to $z=30$ (dark).}
    \label{fwedge}
\end{figure}

Another approach is to isolate the foregrounds rather than subtract them, as described in the introduction. This technique uses the 2D power spectrum in \kpar - \kperp\ space to identify and remove a foreground `wedge'. Although we have made substantial progress in the last year on calibrating the full international LOFAR telescope, there is not yet an appropriate data set with which to start investigating the foreground wedge specifically for large wavenumbers using LBA data. We expect this situation to change within the next year, but here we show the theoretical foreground wedge (Eqn. 5 in \cite{chapman}) in Figure~\ref{fwedge}, for the international stations of LOFAR (the green band in the left-hand panel of Figure~\ref{fig1}). The ranges of \kperp\ and \kpar\ which LOFAR will probe are redshift dependent, although it is clear from Figure~\ref{fwedge} that this does not change significantly within the allowable bandwidth of the LBA. The theoretical foreground wedge is shown, calculated for the specific redshift examples in the Figure, and assuming a 5 deg$^2$ field of view. This is the field of view of a single LOFAR pointing which has been calibrated including the international stations. Sources outside this field of view are typically subtracted, but residual power may still leak in from imperfect source subtraction. We therefore also plot the `worst case' wedge limit as dotted lines, assuming the full 20 deg$^{2}$ field of view of LOFAR calibrated with only the Dutch stations.

The wedge will clearly limit the range of \kpar\ values. In the best case scenario (no foreground leakage; solid lines in Figure~\ref{fwedge}), we can reach the full range of \kperp\ values by setting a lower limit of \kpar$\gtrsim1.8\,$\permpc\ which corresponds to a maximum allowable bandwidth of $\sim$98$\,$kHz. 

\section{Sensitivity predictions}
From the proceeding two sections, we see that we are limited to a narrow range in \kperp\ and \kpar\ that depends on the foreground wedge. This translates to a limitation in baseline length and bandwidth. In this section, we will calculate the sensitivity of LOFAR to these angular and spectral scales. While we do not expect this to be sufficient for a detection of the 21 cm absorption from the Dark Ages, it is instructive to see how far off we are with this particular array design. 

The sensitivity of an an array depends on many factors. In general, the sensitivity for an interferometer is defined as:
\begin{equation}
\label{eqn:sens}
    \Delta S = \frac{S_{\textrm{sys}}}{\sqrt{N(N-1)\delta \nu \delta t}}
\end{equation}
where $S_{\textrm{sys}}$ is the System Equivalent Flux Density (SEFD), $N$ is the number of stations, and $\delta \nu$, $\delta t$ are the bandwidth and integration time, respectively. This equation is valid if all elements of the interferometer are the same, which is not the case for LOFAR. We must account for the different SEFDs (related to the collecting area) of the core, remote, and international stations, which are all different. Expanding Equation~\ref{eqn:sens} for LOFAR, we get:
\begin{equation}
\label{eqn:array}
    \Delta S = \frac{W}{\sqrt{2(2\delta \nu \delta t)\left[ \frac{N_{c}(N_{c}-1)}{2S_{c}^2} + \frac{N_{c}N_{r}}{S_{c}S_{r}} + \frac{N_{c}N_{i}}{S_{c}S_{i}} + \frac{N_{r}(N_{r}-1)}{2S_{r}^2} + \frac{N_{r}N_{i}}{S_{r}S_{i}} + \frac{N_{i}(N_{i}-1)}{2S_{i}^2} \right] }}
\end{equation}
where $W$ is the imaging weight (assumed here to be unity), $\delta \nu$, $\delta t$, and $N$ are as above, with the subscripts $c$, $r$, and $i$ denoting core, remote, and international stations respectively. The system sensitivity, $S$, depends on the effective collecting area of each station and the system temperature:
\begin{equation}
    S_{\textrm{sys}} = \frac{2 \eta k_B}{A_{\textrm{eff}}T_{\textrm{sys}}}
\end{equation}
where $\eta$ is an efficiency factor (assumed here to be unity), $k_B$ is the Boltzmann constant, $A_{\textrm{eff}}$ is the effective collecting area of a station, and the system temperature $T_{\textrm{sys}}$ is the summation of the sky temperature $T_{\textrm{sky}}$ and the instrument temperature, $T_{\textrm{sys}}$. We use tabulated values of $T_{\textrm{sys}}$ for each type of station, and $T_{\textrm{sky}} = 60 \lambda^{2.55}$ to calculate the senstivity of each type of station. This follows the methods of \cite{nijboer} should the reader desire a more in-depth discussion. 

\begin{figure}[!h]
    \centering
    \includegraphics[width=2.5in]{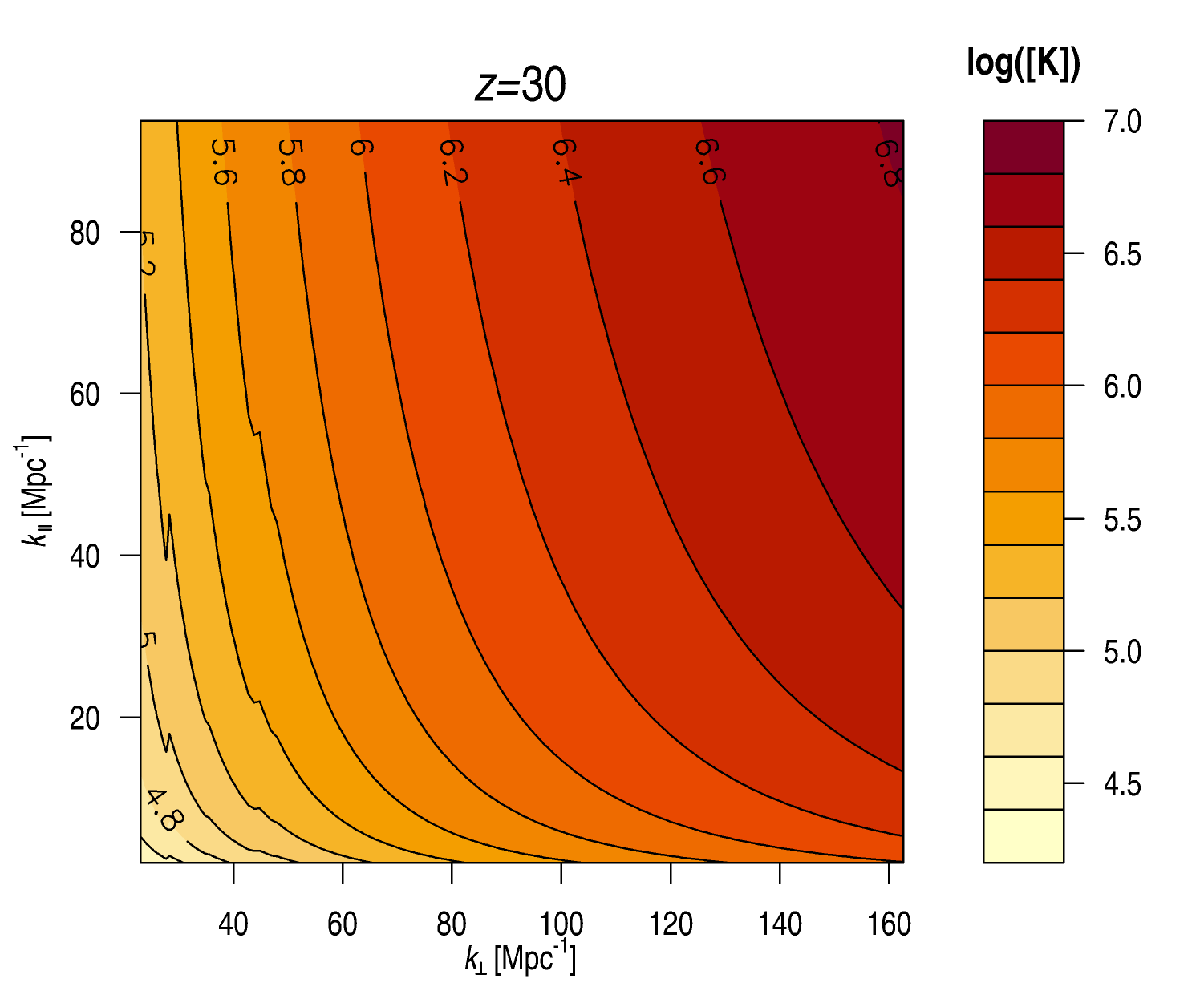} \includegraphics[width=2.5in]{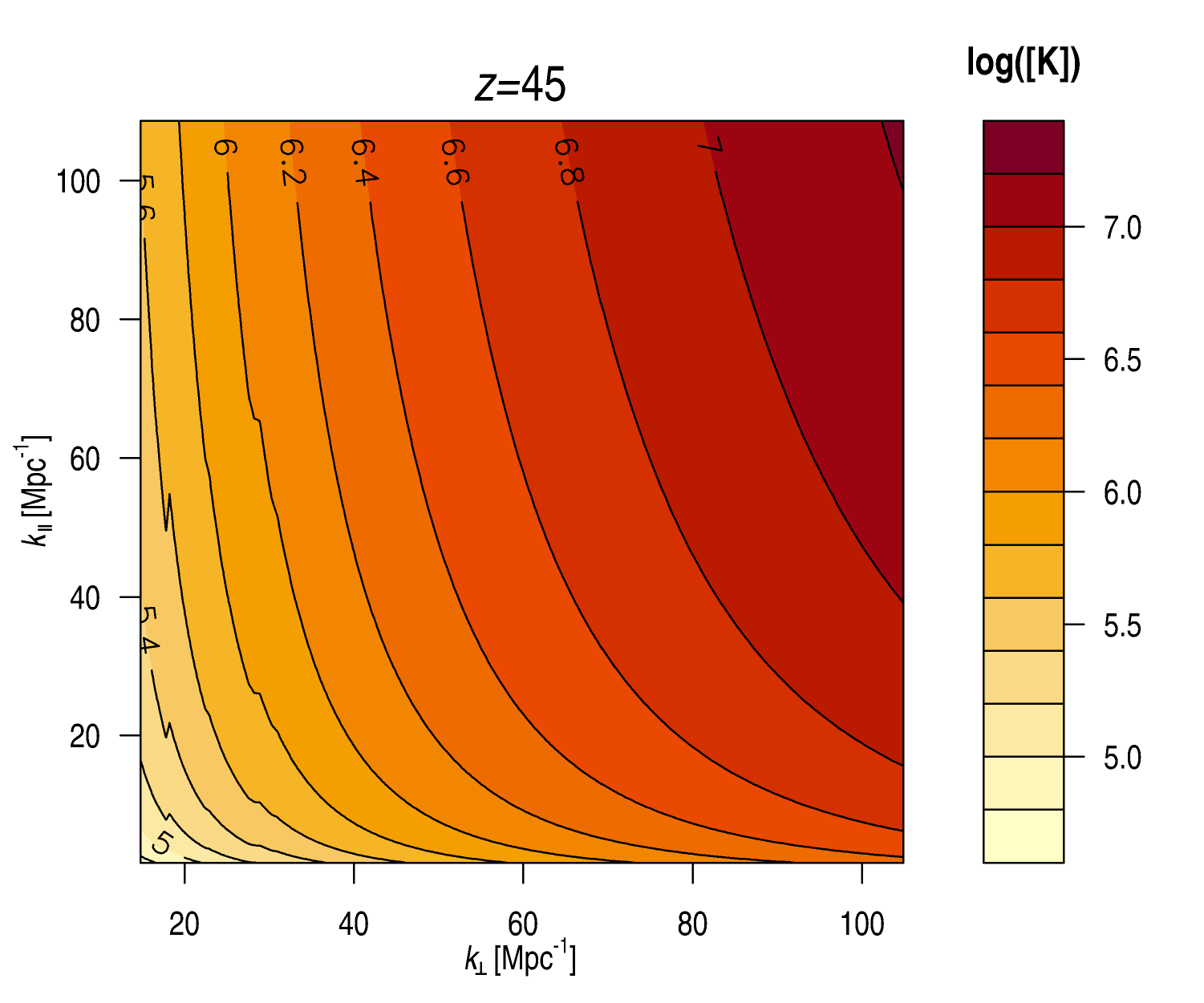}\\
    \includegraphics[width=2.5in]{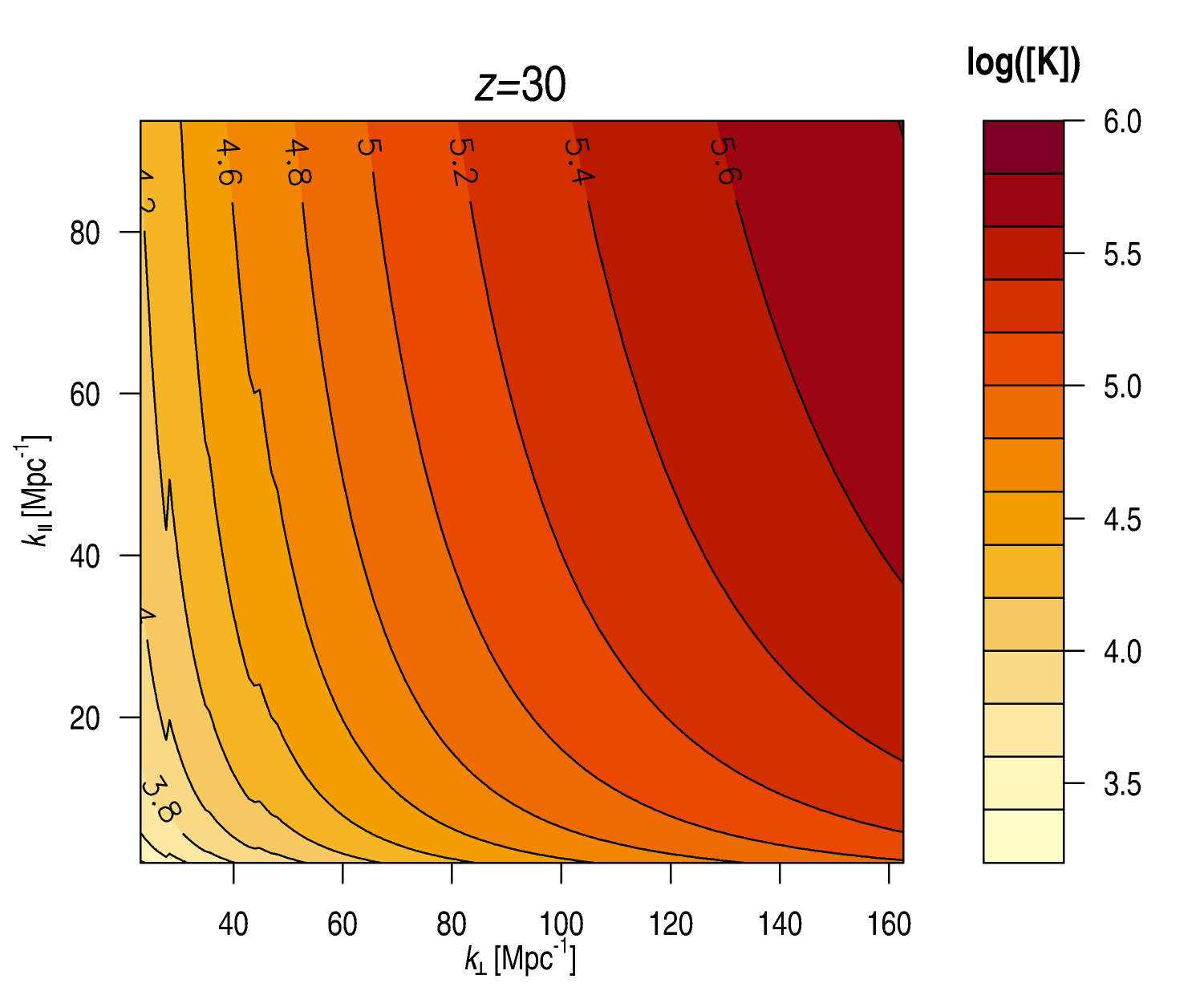}    
    \includegraphics[width=2.5in]{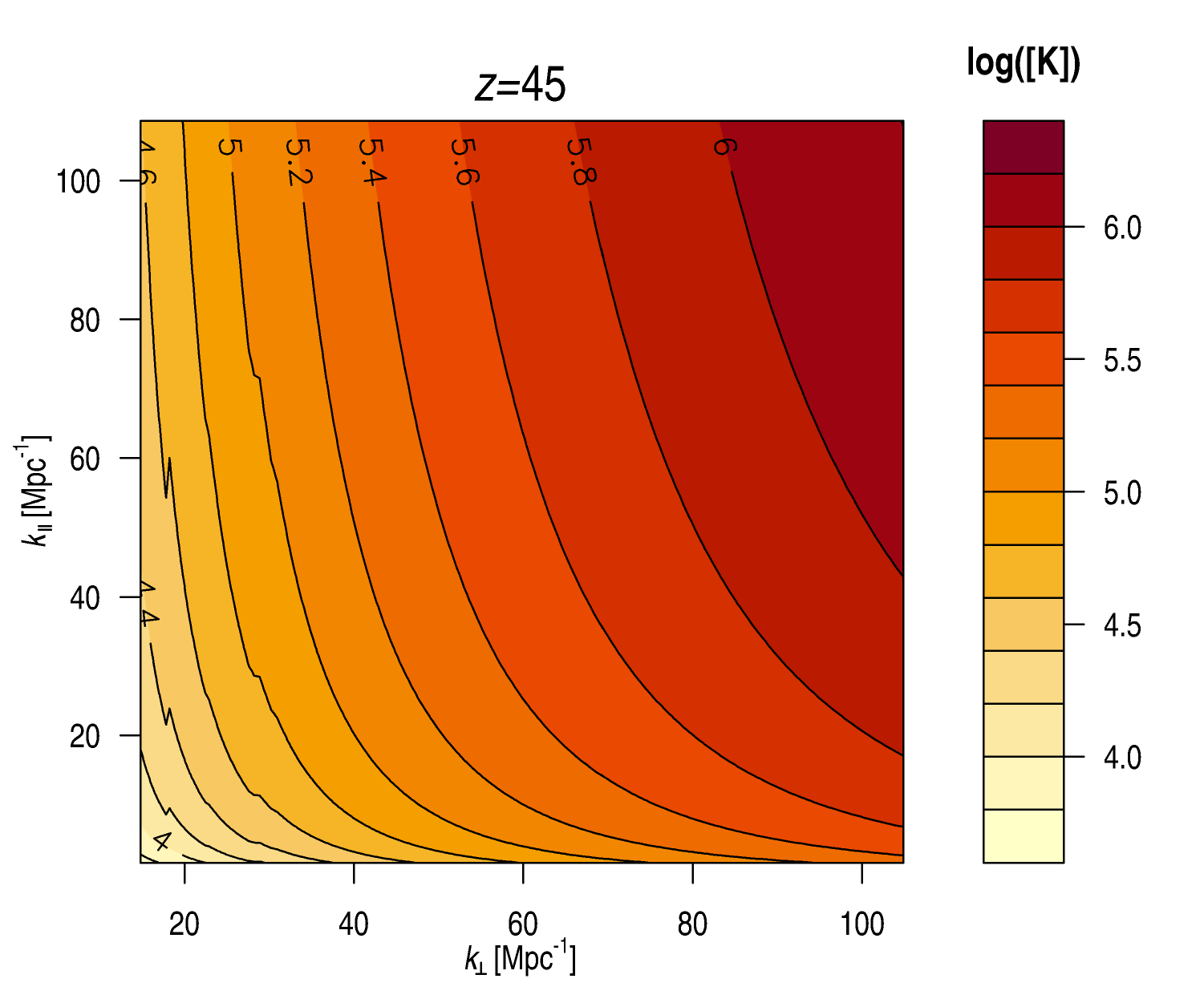}
    \caption{The log of the temperature brightness to which LOFAR is sensitive, as a function of \kperp\ and \kpar\ at $z=30$ (left column) and $z=45$ (right column) for an 800 hour observation (top row) and a 10 year observation (bottom row).}
    \label{sens}
\end{figure}

Here we adapt the sensitivity calculations to the specific idea we want to investigate. For a given redshift, \kperp\ and  \kpar\ can be translated into values of spatial resolution and bandwidth. We use the range of the spatial resolutions to specify which LOFAR stations will contribute to the sensitivity on those spatial scales, and only these are used in Equation~\ref{eqn:array}. Note that the shorter baselines will still contribute to the overall sensitivity at smaller scales: a perfect point source will be detected on both short and long baselines. The converse is not true -- if a source is extended, it may be detected on short baselines but not long baselines. Once the stations are selected, we calculate the array sensitivity as in Equation~\ref{eqn:array} using only those stations. Finally, we convert the sensitivity to brightness temperature in Kelvin: 
\begin{equation}
    T_b = \frac{S c^2 4 ln(2)}{2 \nu^2 k_B \pi \theta^2}
\end{equation}
where $S$ is the array sensitivity from Equation~\ref{eqn:array}, $c$ is the speed of light, $\nu$ is the observing frequency, $k_B$ is the Boltzmann constant, and $\theta$ is the FWHM of a circular resolution element. From this we can see that $T_b$ depends inversely on the resolution, and thus for smaller resolution elements (higher \kperp\ values) we are sensitive to higher values of $T_b$. It is useful to keep in mind that Equation~\ref{eqn:array} depends on the individual stations' effective collecting area, the bandwidth, and the integration time. An example of the sensitivity for $z=30$ and $z=45$ is given in Figure~\ref{sens}, for observation times of 800 hours and 10 years. Values of $T_b$ for four different redshift slices are given in Table~\ref{t1}, for the 10 year observation. 

\begin{table}[!h]
\caption{$T_b$ sensitivity values for 10 year observation time, for four different redshift slices.}
\label{t1}
\begin{tabular}{cccccccc}
\hline
$z$ & \kperp\ range & $\theta$ range & \kpar\ range & $\Delta\nu$ range & $T_{b,\textrm{min}}$ & $T_{b,\textrm{max}}$ & $T_{b,\textrm{median}}$ \\ 
& [\permpc] & [arcsec] & [\permpc] & [kHz] & [K] & [K] & [K] \\ 
 \hline \\[-8pt] 
30 & $10^{1.4}$ - $10^{2.2}$ & 5.9 - 0.8 & $10^{0.3}$ - $10^{2.0}$ & 100.0 - 2.1 & $10^{3.4}$ & $10^{5.8}$ & $10^{4.5}$ \\ 
35 & $10^{1.3}$ - $10^{2.1}$ & 6.9 - 1.0 & $10^{0.3}$ - $10^{2.0}$ & 100.3 - 1.9 & $10^{3.5}$ & $10^{6.0}$ & $10^{4.7}$ \\ 
40 & $10^{1.2}$ - $10^{2.1}$ & 7.8 - 1.1 & $10^{0.2}$ - $10^{2.0}$ & 100.2 - 1.7 & $10^{3.6}$ & $10^{6.1}$ & $10^{4.8}$ \\ 
45 & $10^{1.2}$ - $10^{2.0}$ & 8.8 - 1.2 & $10^{0.2}$ - $10^{2.0}$ & 100.1 - 1.5 & $10^{3.7}$ & $10^{6.2}$ & $10^{4.9}$ \\ 
\hline 
\end{tabular}
\vspace*{-4pt}
\end{table}

\section{Discussions}

The 21 cm signal from the Dark Ages is expected to be $\lesssim 200\,$mK \cite{pritchard}, but could be as small as $\sim 1\,$mK at the most pessimistic, in the standard cosmology. Assuming an optimistic case of $\sim 100\,$mK, with 10 years of observations and assuming no data loss or corruption from calibration, Figure~\ref{sens} indicates the best LOFAR can do is $\sim10^4\,$K. Stacking in redshift slices can help alleviate this: the most one can stack is the entire 48$\,$MHz bandwidth. For the limited allowed values of bandwidth, this would increase the signal to noise by a factor in the range of $20$ - $400$. However, if the goal is to detect individual sources in absorption to construct the 3-point correlation function, stacking in redshift will not help. We therefore conclude that we are approximately five orders of magnitude away from the desired sensitivity.

As mentioned in the introduction, this signal could be boosted by primordial black holes, which could narrow the gap between expected and achievable sensitivity by 2 - 3 orders of magnitude depending on the number and combined Eddington ratio of primordial black holes. However, other scenarios have also been proposed. The EDGES absorption profile \cite{bowman}, although not yet independently confirmed, is more than twice as deep as expected from standard cosmology. This necessitates either an increase in the radio background, or cooling in the IGM (or both), which would also impact the Dark Ages signal (see, e.g. Fig. 1 in \cite{furlanetto} for a demonstration of models invoking extra cooling). New exotic physics have been proposed to explain the gap between the EDGES detection and standard cosmology. For example, baryon-dark matter scattering can increase the global signal and its r.m.s fluctuations at the relevant Dark Ages redshifts by over an order of magnitude \cite{fialkov,munoz_b}. Another option for cooling baryons is for a small fraction of dark matter to have a mini-charge \cite{munoz_c,berlin} which allows coupling with photons. More exotic options for explaining the stronger than expected EDGES detection include (but are not limited to): interacting dark energy \cite{costa}, axions \cite{moroi,pospelov}, and neutrino decay \cite{chianese}. Not all of these predict significant increases for the redshift ranges consider here, and thus any detection would help place constraints on these models.

\section{Conclusion and Future Work}

In this paper we have considered the question of probing the small-scale 21 cm power spectrum from the Dark Ages, using LOFAR, which is the only currently operational telescope capable of reaching sub-arcsecond scales at the relevant low frequencies. 
This resolution is interesting for probing high values of \kperp , which can help place new constraints on models of inflation. Specifically, it is interesting to test whether the scalar spectral index measured by Plank for $k \lesssim 2\,$\permpc\ holds at higher values of $k$ or if it there are deviations from primordial non-Gaussianity, as discussed in \cite{joudaki,mao,chongchitnan}. This can have far-reaching implications, as an amplification of power in the smaller angular scales (higher $k$) is necessary for the formation of primordial black holes of sufficient size to help explain the presence of super-massive black holes in galaxies at high redshifts. 

We have shown that the LOFAR LBA is sensitive to $k_{\perp} \sim 10^{ 1.5 }$ - $10^{ 2.3 }\,$\permpc\ based on the international stations which provide its angular resolution; and $k_{\parallel} \sim 10^{ -2.4 }$ - $10^{ 2.4 }\,$\permpc\ based on its flexible spectral resolution. These values of \kperp\ and \kpar\ overlap with the theoretical expectation of where the `foreground wedge' is located, which places a lower limit on \kpar\ of $\sim10^1\,$\permpc . Using these ranges of \kperp\ and \kpar , we have estimated the LOFAR sensitivity in terms of temperature brightness for four redshift slices accessible from the LBA frequency range, for 10 years of integration time. We find that we can reach $\sim 10^4\,$K at best without stacking in redshift to improve signal to noise, in the absence of any calibration errors or incomplete foreground isolation. This is five orders of magnitude above the expected strength of the 21 cm signal from the Dark Ages, and LOFAR clearly does not have sufficient collecting area. 

The SKA-LOW is expected to be an order of magnitude more sensitive than LOFAR, but the design baseline does not include the long baselines to match LOFAR's unique spatial resolution, and it will not extend to frequencies below 50 MHz. Although LOFAR cannot achieve the desired sensitivity, it is useful as a testbed to quantify the foreground isolation technique for high values of \kperp\  by testing this on real data. At the time of writing, no fully calibrated LBA dataset with the international baselines exists, but we expect this to change within the next year. 

Further work should include an assessment of improving the collecting area of a low-frequency array with long baselines. For example, when designing a lunar array, the sensitivity of the individual dipoles must be improved over current existing hardware, then their station configuration must optimised to increase collecting area, and finally the array configuration must be taken into account to maximise the sensitivity to specific angular scales. Investigating the foreground wedge at high values of \kpar , i.e., for long baselines, will be most useful for optimising the array configuration.

\vskip6pt

\enlargethispage{20pt}

\dataccess{This analysis was carried out using R and all scripts are available at \url{https://github.com/lmorabit/cosmic_dawn}.}

\aucontribute{LKM carried out the calculations and authored the manuscript. JS conceived the project and provided guidance for relevant content. }

\competing{The authors declare that they have no competing interests. }


\ack{The authors would like to thank the anonymous referees for their helpful comments. The authors acknowledge useful conversations with Ian Harrison. LKM gratefully acknowledges the instructive and well-commented code from \url{https://gitlab.com/radio-fisher}, which is described in \cite{bull}. }


\end{document}